# Supersonic radiatively cooled rotating flows and jets in the laboratory

D.J. Ampleford[1,*], S.V. Lebedev[2], A. Ciardi[3], S.N. Bland[2], S.C. Bott[2], G.N. Hall[2], N. Naz[2],

C.A. Jennings[1], M. Sherlock[4], J.P. Chittenden[2], J.B.A.Palmer[2], A. Frank[5], E. Blackman[5]

[1] *Sandia National Laboratories, Albuquerque, NM 87123-1106, USA*

[2] *Blackett Laboratory, Imperial College, London SW7 2BW, United Kingdom*

[3] *Observatoire de Paris, LERMA, Meudon, 92195, France*

[4] *Rutherford Appleton Laboratory, Didcot, OX11 0QX, United Kindom*

[5] *Department of Physics and Astronomy, and Laboratory for Laser Energetics, University of Rochester, Rochester NY, USA*

PACS:   98.38.Fs. 47.40.Ki, 52.58.Lq

**Abstract:**

The first laboratory astrophysics experiments to produce a radiatively cooled plasma jet with dynamically significant angular momentum are discussed. A new configuration of wire array z-pinch, the twisted conical wire array, is used to produce convergent plasma flows each rotating about the central axis. Collision of the flows produces a standing shock and jet that each have supersonic azimuthal velocities. By varying the twist angle of the array, the rotation velocity of the system can be controlled, with jet rotation velocities reaching ~20% of the propagation velocity.

Collimated jets and winds that emerge from magnetized accretion star-discs systems are partly responsible for the extraction of the angular momentum from the accreting matter [1] and play a vital role in the star formation process. Their formation and kinematics have been studied both analytically and in multi-dimensional numerical simulations [1] and a certain level of angular momentum is expected to be present in the ejected flows. Observation of rotation in proto-stellar jets has also been reported [2]. In addition to the astrophysical relevance, rotation has also been suggested as a means to stabilize the Rayleigh-Taylor and MHD instabilities in z-pinch implosions [3], with possible application to Inertial Confinement Fusion.

Complementary to the 'traditional' astrophysical studies, experiments have shown the capability to reproduce in the laboratory complex, scaled plasma dynamics (see [4] for a recent review) providing insights into the behaviour of astrophysical phenomena and offering benchmark data for the numerical models. Previously reported experimental studies of supersonic jets, both hydrodynamic [5,6,7] and magnetically driven [8,9], dealt with non-rotating jets. The controllable introduction of rotation into High Energy Density plasma experiment is, however, a non-trivial task and has not previously been reported in the literature. This Letter reports on the first laboratory experiments to form rotating supersonic plasma jets.

An overview of the experiments is shown in Fig.1, which consists of a conical array of fine metallic wires (16 x 18μm W) driven by a 1MA, 250ns current pulse on the Magpie facility. The array is 10mm high, the diameter at the base is 16mm with an inclination angle with respect to the axis of $\beta = 30°$. In a few nanoseconds resistive heating of the wires converts them into a heterogeneous structure consisting of a cold dense core surrounded by a hot (~ 20 eV) low density coronal plasma [10]. The plasma is continuously ablated from the wire cores and is accelerated by the JxB Lorentz force towards the axis, creating a quasi-stationary supersonic converging conical flow. We note that previous experiments have demonstrated that negligible current is advected to the array axis by these streams [11]. The stagnation on axis produces a standing shock which effectively re-directs the flow in the axial direction, producing a supersonic jet. The typical dimensionless parameters of the laboratory jets [6] are similar to those of the jets observed in young stars: Mach number (ratio of jet velocity to sound speed) $M \sim 20$; the cooling parameter (ratio of cooling length to characteristic scale length) $\chi \leq 1$; Reynolds number (ratio of inertial to viscous forces) Re>>1 and Peclet number (ratio of heat convection to heat diffusion) Pe>> 1. In the present experiments the top of the array is rotated with respect to the bottom electrode so that the wires are twisted about the axis of the array by a twist angle $\Phi$ (see Fig. 1b and Fig. 2d). The bulk of the current flows in the close vicinity of the wire cores and an axial magnetic field is generated. Divergence of this field, along with Eddy currents in the electrodes lead to a significant radial magnetic field. Both of these components contribute to an azimuthal component of the Lorentz force. The ablated plasma is then accelerated off the wires with a non-zero angular momentum producing a supersonic rotating converging plasma flow. The magnitude of the axial and radial magnetic fields, the associated azimuthal force and resulting angular momentum can be controlled by varying $\Phi$.

The images in Fig. 1 show the diameters of both the standing shock and the jet increase with increasing $\Phi$ and the resulting jets become more divergent. End on imaging at early times (t~210ns, Fig 2), before the jet formation, demonstrates the effects of angular momentum on the standing shock. The XUV images ($h\nu > 30$ eV) show bright emission coming from both the wires and the converging plasma flows. Its intensity rapidly decays with distance from the wires, indicative of strong radiative cooling in the flow. However, bright emission is observed again on the arrays axis where the kinetic energy of the flow is thermalized at the standing shock. For the more energetic spectral range $h\nu \sim 200-290$ eV (bottom row in Fig 2) only the shocked plasma is observed. For the $\Phi \neq 0$ cases, the emission peaks off axis, clearly indicating a hollow radial emission profile. The measured shock diameter as a function of twist angle (shown in Fig. 3) increases for larger $\Phi$. The absence of emission from the central region of the standing shock and the dependence of the size of the shock on the initial array twist angles $\Phi$ are both consistent with the presence of angular momentum in the converging plasma flow preventing the complete stagnation of the plasma onto the array axis. Experiments with twisted cylindrical wire arrays have demonstrated that the magnetic field on the axis of the system is incapable of preventing collapse of the column onto the axis, indicating that in the present experiments it is the angular momentum and not the magnetic field which is responsible for the observed changes in the structure of the standing shock.

The equilibrium radius of the standing shock can be determined by assuming the centrifugal force arising from rotation of the plasma in the standing shock is balanced by the ram pressure of the incoming plasma flow. Force balance in the shock can be written as

$$m \frac{V_\phi^2}{r_{shock}} = \frac{dm}{dt} V_{flow} \cos\beta \sin\gamma, \qquad (1)$$

where $\gamma$ is the angle between the flow and the tangential direction on the shock surface (see fig 3b) and $m$ is mass accumulated in the rotating shock. Assuming that there is no torque present in the flow, the accumulated angular momentum in the standing shock is equal to the total injected angular momentum:

$$m V_\phi r_{shock} = m R_0 V_{flow} \cos\beta \sin\alpha \qquad (2)$$

$\alpha$ is the flow angle at the point of injection. Analogous to the ablation phase in cylindrical wire array Z-pinches [10] we assume that the majority of the current remains in close vicinity of the wires, then the plasma ablation is characterized by a rocket model [10], where the flow inside the array has a constant velocity $V_{flow} = 150$ km s$^{-1}$, and equations 1 and 2, can be solved to give the equilibrium shock radius as a function of time and of flow injection angle $\alpha$. However $\alpha$ cannot be measured accurately and we take it as a free parameter such that the solution fits the equilibrium radius $r_{shock} = 1.1$ mm for a twist angle $\Phi = 2\pi/32$, which requires $\alpha = 3.3°$. For the other twist angles the ratio $\alpha/\Phi$ is taken as constant, which is consistent with an azimuthal force $F_\phi \propto \phi$. The solution for $t = 210$ ns is plotted in Fig. 3a. The good fit to all experimental data suggests that the simple analytical model correctly describes the dependence of the shock radius on the twist angle. The inferred values of $\alpha$ are consistent with those calculated from a Biot-Savart magneto-static field solver and in particular with the 3D resistive MHD simulations discussed later. The rotation velocities of the plasma in the standing shock, suggested by the model, are 23, 35 and 53 km s$^{-1}$ for the three increasing twist angles. From the estimated maximum temperature ($T_e < 50$ eV) and average charge state ($\bar{Z} < 5$) in the shock [10] we find that the rotation velocity is larger that the local sound speed $c_s < 15$ km s$^{-1}$ and the flow in the standing shock is supersonically rotating. For the array with $\Phi = 2\pi/16$ the rotational Mach number is > 3.

Imaging flow trajectories in hot plasma is non-trivial. One possibility is that when small perturbations are present in the plasma (either imposed or naturally occurring) these can be used as tracers of the evolution of the system. A typical jet produced in a twisted conical array ($\Phi = 2\pi/64$) is shown in Fig 4a. In this shadowgraph image, which is sensitive to electron density gradients, two filamentary structures in the shape of a double-helix are visible. Interferometer images of the jet, such as the upper part of Fig 1b, show a cylindrically symmetric, quasi-hollow, electron density distribution. In the interferometer image a filament is visible, however analysis of this image indicates this only constitutes a ~30% increase in electron density at this isolated azimuthal location. As this perturbation is over a small fraction of the total circumference of the jet (~0.2mm of a hollow 3mm diameter jet) the filament only accounts for a small (few %) fraction of the mass, so is unlikely to be dynamically significant. The sharp gradients present within these filaments enable them to be imaged by shadowgraphy. In the experiments the filaments are also observed in the other cases (Fig. 4b-d); however for the jets produced in zero-twist conical arrays, the filaments show no rotation (Fig 4b). Such structures are also observed in magnetically driven laboratory jets produced on radial wire arrays [8,12] and are not a specific feature of conical wire arrays. The filaments can in fact be linked to the discrete nature of the convergent plasma streams which imprint density modulations on the shock and jet. Assuming the rotation of the jet produces the helical structure, the filaments can be used to estimate the ratio of the axial $V_z$ to the rotation velocity $V_\phi$ in the jet. Where these two filaments cross, at the centre of the image (marked point 'A' on Fig 4a), the *pitch* angle is $\sim 10°$ giving a ratio $V_\phi/V_z \sim 0.18$. Taking the axial flow velocity at this point to be equal to the jet tip velocity $200 \pm 50$ km s$^{-1}$ [5,13] gives an upper estimate for the rotation velocity $V_\phi = 35 \pm 9$ km s$^{-1}$. However, the results of numerical simulations [13] show that the flow velocity in the body of the jet is not constant but increases with distance from the top of the shock. A jet velocity of ~ 130 km s$^{-1}$, consistent with simulations [13], is required to give a jet rotation velocity equal to the shock rotation velocity (23 km s$^{-1}$) inferred from the model. Similar *braiding* of jets was observed in the extragalactic jet NGC4258 [14] and has been used as a diagnostic for rotation velocities.

Figure 4b-d shows that for increasing angular momentum the jet becomes less collimated and highly divergent: as expected angular momentum has a detrimental effect on the jet collimation. Electron density profiles of the jets, obtained by Abel inversion of the phase shifts on interferometer images show a somewhat hollow mass distribution, which is in agreement with the end-on self-emission imaging of the conical shock discussed earlier (see for example Fig 2).

Numerical modelling of the experiments was performed with the 3D resistive MHD code GORGON [12,15]. A typical simulation result for an array $\Phi = 2\pi/32$ at $t = 270$ ns is shown in Fig. 5a. The isodensity surfaces show both the plasma streams ($\rho = 2\times10^{-5}$ g cm$^{-3}$) and the dense plasma around the wires ($\rho = 10^{-3}$ g cm$^{-3}$). The streams merge close to the axis, where they form a standing shock which redirects the plasma axially into a jet. The emerging jets are radiatively cooled and hypersonic ($M \sim 10-20$), with typical velocities $V_z \sim 100-200$ km s$^{-1}$, electron temperatures $T_e \sim 15-20$ eV and densities $\rho \sim 10^{-5}-10^{-6}$ g cm$^{-3}$. The axial profiles are similar to those seen in conical arrays [13], with decreasing density and increasing axial velocity with height. From the MHD simulations we find that the relative importance of the two azimuthal components of the Lorentz force vary along the length of the wire: the $j_z B_r$ dominates near the electrodes while the $j_r B_z$ is stronger over the central region of the wires. Poloidal magnetic field lines (Fig5a) show the radial magnetic field $B_r$ reversing sign along the wire, which corresponds to a change of direction of the $j_z B_r$ component of the Lorentz force. However the total Lorentz force, which depends also on the current density distribution, changes sign and generates a counter-rotating flow only in the region close to the top electrode (~ 2 mm). Velocity streamlines from one of the wires clearly indicate the presence of rotation in both the shock and jet ($V_\phi/V_z = 0.1-0.3$); the uppermost streamlines show the region where the flow reverses the sense of rotation. There, the low rate of mass ablation and the increased time-of-flight needed to reach the axis produce relatively low density plasma streams which do not affect the jet rotation profile. The azimuthally averaged profiles of density, temperature and azimuthal velocity are shown in Fig 5b at 200 ns for the three twist angles. Typical rotation velocities are consistent with the estimates obtained from the analytical model and in the simulations the post-shock flow is supersonically rotating ($M \sim 2-3$). The density peaks off axis, while it sharply drops inside the "hollow" shock. Typical temperatures in the shocked plasma are $T_e \sim 40$ eV. The radius of the shock determined from the position of the density peak for each twist angle is plotted in Fig. 3 and shows excellent agreement with the experimental data and the analytical model. Because of the finite length ($\sim 3$ mm) over which the plasma streams are redirected and accelerated there is no single flow injection angle $\alpha$ at the wires. However an estimate of $\alpha$ can be obtained from the flow angle determined inside the array; the values obtained in the simulations are $\alpha \sim 1.4°$, $2.0°$ and $4.1°$ for the arrays with twist angles $\Phi = 2\pi/64$, $2\pi/32$ and $2\pi/16$ respectively. The analytical model predicts slightly larger angles $\alpha$, thus overestimating the injection of angular momentum at the wires. The simulations show that due to non-zero magnitude of current inside of the array, the angular momentum increases with radius by a factor $\sim 2$. At the shock front, a radial force balance exists between the centrifugal force and the ram pressure, and the magnetic fields do not play any significant dynamical role in the shock and jet. The experiments are most relevant to the propagation regime of proto-stellar jets, where thermal and kinetic effects dominate the dynamics [16].

In summary, experiments have been performed to produce rotating convergent flows, standing shocks and hypersonic jets in the laboratory. By varying the twist angle $\Phi$ of the array, the amount of angular momentum that is introduced in the system can be controlled and it shows a detrimental effect on the collimation of the jets when no ambient medium is present. The properties of the rotating standing shock can be well described by an analytical force balance model. MHD simulations reproduce many of the features of the experiments and provide insights into the dynamics of the system and applicability to proto-stellar jets. The rotation in the jet found by experimental measurements and simulations indicate that the ratio of $V_\phi/V_z \sim 0.18$, together with the Peclet and Reynolds numbers, and the cooling parameter are in the same range as those of protostellar jets. The experiments discussed can be adapted to study the propagation dynamics and interactions in an ambient medium [10], or to combine with other pulsed-power driven jet experiments [8,12] to study the effect of rotation on magnetically driven jets. Both types of experiment are planned for the future.

**Figures:**

**FIG 1:** Laser interferometry of (a) an untwisted conical wire array and (b) a conical wire array with a twist between the two electrodes $\Phi=2\pi/16$. The twist is seen to alter the conical shock on the axis of the array and the jet which propagates upwards, above the anode plate. The conical arrays consist of 16, 18µm W wires with 30º wire inclination angles.

**FIG 2:** Time gated XUV (top, hv>30eV) and soft x-ray (bottom, hv>220eV) self emission images taken end-on to the array at ~210ns after the start of current. This rotation angle of the array $\Phi$ (defined on image (d)) is varied on the different images: (a) $\Phi = 0$, (b) $\Phi = 2\pi/64$ (c) $\Phi = 2\pi/32$ and (d) $\Phi = 2\pi/16$.

**FIG 3 [Color online]:** Measurements of the conical shock diameter from experiments are plotted (a) at 210ns (squares), with the limits indicating the thickness of the wall of the shock. The line on the plot represents the predictions of the shock model described in the text using the parameters defined in (b). The axis along the top of the plot is the angle of the streams used to fit the model to the data. Triangles on the plot show the position of peak density from MHD simulations described later in the text.

**FIG 4:** Schlieren images of jets from conical wire arrays. (a) shows filament like structures, which rotate around the jet for a twist angle of $\Phi = 2\pi/16$. Also shown are schlieren images of the base of jets with (b) $\Phi = 0$, (c) $\Phi = 2\pi/64$ and (d) $\Phi = 2\pi/16$. All images are at t=315-340ns after start of current.

**FIG 5 [Color]:** Results of MHD simulations. (a) Isodensity surfaces at 270 ns show the dense plasma around the wires ($10^{-3}$ g/cm$^3$, (dark gray) and plasma streams ($2.5 \times 10^{-5}$ g/cm$^3$, light grey); streamlines from one of the wires are shown with the oppositely rotating flows separated visually by red and orange streamlines; the azimuthally averaged poloidal magnetic filed lines are shown in blue; (b) Azimuthally averaged profiles of azimuthal velocity, density and temperature taken 6mm above the cathode at 200ns.

**Figure 1**, Ampleford et al.

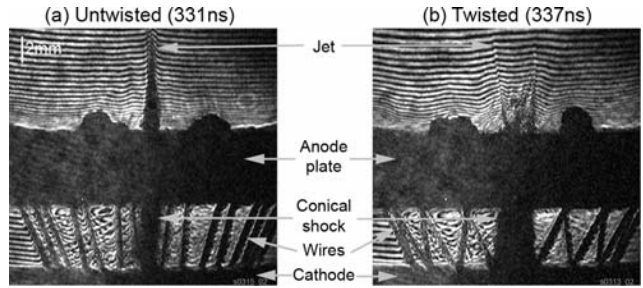

**Figure 2**, Ampleford et al.

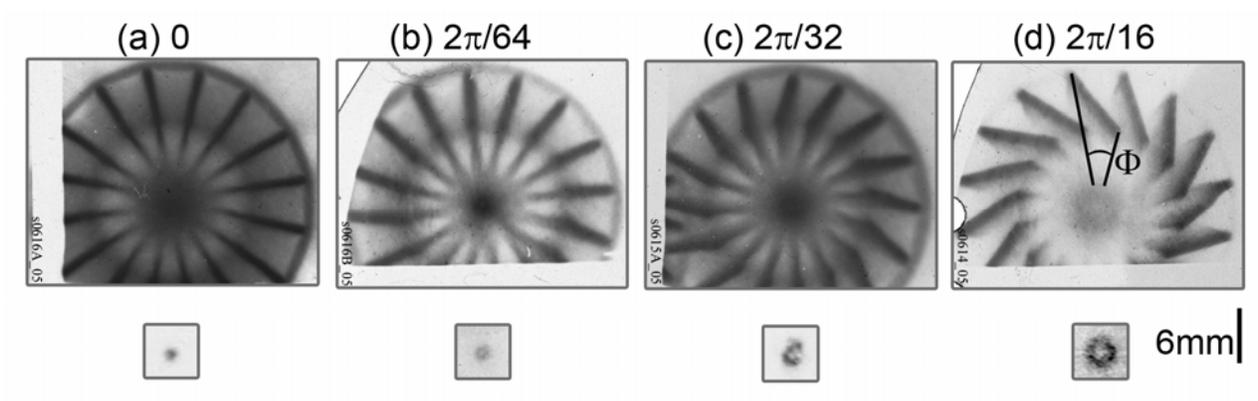

**Figure 3**, Ampleford et al.

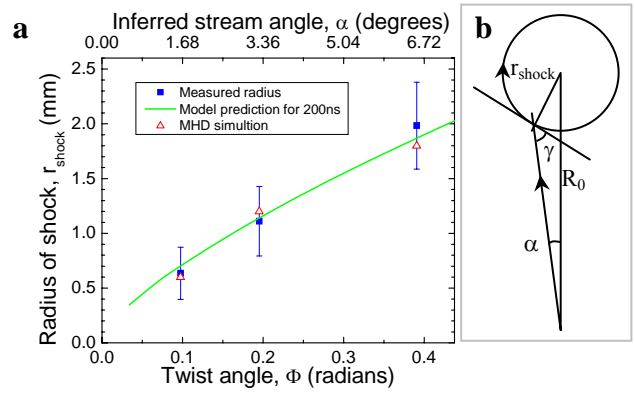

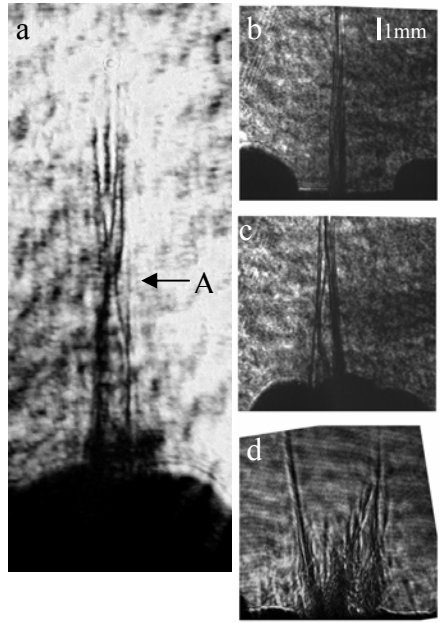

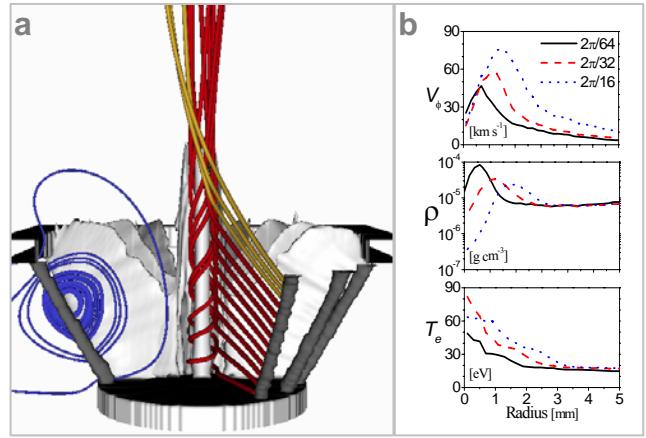